\documentclass[journal = aamick, manuscript = article, layout = twocolumn]{achemso}
\usepackage{graphicx} 
\usepackage{epstopdf} 
\usepackage{dcolumn}
\usepackage{amssymb}
\usepackage{setspace}
\usepackage{balance}
\usepackage{array} 
\usepackage{natbib}
\usepackage{amsmath}
\usepackage{hyperref}
\DeclareMathAlphabet{\mathpzc}{OT1}{pzc}{m}{it}

\usepackage[usenames,dvipsnames,table,xcdraw]{xcolor}

{\title{Photoacoustic Sensing of Trapped Fluids in Nanoporous Thin Films: Device Engineering and Sensing Scheme}

\author{Giulio Benetti}
\affiliation{Interdisciplinary Laboratories for Advanced Materials Physics (I-LAMP), Universit\`a Cattolica del Sacro Cuore, Via Musei 41, 25121 Brescia, Italy}
\alsoaffiliation{Dipartimento di Matematica e Fisica, Universit\`a Cattolica del Sacro Cuore, Via Musei 41, 25121 Brescia, Italy}
\alsoaffiliation{Laboratory of Solid State Physics and Magnetism, Department of Physics and Astronomy, KU Leuven, Celestijnenlaan 200D, B-3001 Leuven, Belgium}

\author{Marco Gandolfi}
\affiliation{Interdisciplinary Laboratories for Advanced Materials Physics (I-LAMP), Universit\`a Cattolica del Sacro Cuore, Via Musei 41, 25121 Brescia, Italy}
\alsoaffiliation{Dipartimento di Matematica e Fisica, Universit\`a Cattolica del Sacro Cuore, Via Musei 41, 25121 Brescia, Italy}
\alsoaffiliation{Laboratory for Soft Matter and Biophysics, Department of Physics and Astronomy, KU Leuven, Celestijnenlaan 200D, B-3001 Leuven, Belgium}

\author{Margriet J Van Bael}
\affiliation{Laboratory of Solid State Physics and Magnetism, Department of Physics and Astronomy, KU Leuven, Celestijnenlaan 200D, B-3001 Leuven, Belgium}

\author{Luca Gavioli}
\affiliation{Interdisciplinary Laboratories for Advanced Materials Physics (I-LAMP), Universit\`a Cattolica del Sacro Cuore, Via Musei 41, 25121 Brescia, Italy}
\alsoaffiliation{Dipartimento di Matematica e Fisica, Universit\`a Cattolica del Sacro Cuore, Via Musei 41, 25121 Brescia, Italy}

\author{Claudio Giannetti}
\affiliation{Interdisciplinary Laboratories for Advanced Materials Physics (I-LAMP), Universit\`a Cattolica del Sacro Cuore, Via Musei 41, 25121 Brescia, Italy}
\alsoaffiliation{Dipartimento di Matematica e Fisica, Universit\`a Cattolica del Sacro Cuore, Via Musei 41, 25121 Brescia, Italy}

\author{Claudia Caddeo}
\affiliation{Istituto Officina dei Materiali (CNR - IOM) Cagliari, Cittadella Universitaria, I-09042 Monserrato (Ca), Italy}

\author{Francesco Banfi}
\email{francesco.banfi@unicatt.it}
\affiliation{Interdisciplinary Laboratories for Advanced Materials Physics (I-LAMP), Universit\`a Cattolica del Sacro Cuore, Via Musei 41, 25121 Brescia, Italy}
\alsoaffiliation{Dipartimento di Matematica e Fisica, Universit\`a Cattolica del Sacro Cuore, Via Musei 41, 25121 Brescia, Italy}


\begin{document}
\begin{abstract}
Accessing fluid infiltration in nanogranular coatings is an outstanding challenge, of relevance for applications ranging from nanomedicine to catalysis. A sensing platform, allowing to quantify the amount of fluid infiltrated in a nanogranular ultrathin coating, with thickness in the 10 to 40 nm range, is here proposed and theoretically investigated by multiscale modelling. The scheme relies on impulsive photoacoustic excitation of hypersonic mechanical breathing modes in engineered gas-phase synthesised nanogranular metallic ultathin films and time-resolved acousto-optical read-out of the breathing modes frequency shift upon liquid infiltration. A superior sensitivity, exceeding 26x10$^{3}$ cm\textsuperscript{2}/g, is predicted upon equivalent areal mass loading of a few ng/mm\textsuperscript{2}.
The capability of the present scheme to discriminate among different infiltration patterns is discussed. The platform is an ideal tool to investigate nanofluidics in granular materials and naturally serves as a distributed nanogetter coating, integrating fluid sensing capabilities. The proposed scheme is readily extendable to other nanoscale and mesoscale porous materials.
\end{abstract}

\section{Introduction}\label{sec:first}
Nanofluidics in nanoporous coatings is an emerging topic at the forefront of nanotechnology. The subject, at the cross-road of physics, material science and engineering, is of relevance both scientifically\cite{bocquet2010,schoch2008}, fluid dynamics at the nanoscale differing significantly from its macroscale counterpart, and from a technological stand point, where applications have been proposed throughout disparate fields, ranging from biotechnology\cite{zhang2016} and nanomedicine\cite{van2007,adiga2009,kholmanov2012nanostructured} to gas storage\cite{Kumar2017,morris2008} and catalysis\cite{Bau2017}.

In this context nanoporous films have been proposed as new platforms to investigate nanofluidics\cite{bocquet2010} and the related wettability issues\cite{ceratti_critical_2015,chen_full_2015}.
As for the technological drive, whatever the application, the exploitation of nanoporous getter coatings ultimately relies on their capability to be infiltrated by a fluid\cite{Mercuri}. Despite the demonstrated versatility and wide-spreading of porous thin films, measurements of their permeability remains an outstanding issue. 

This issue is primarily tackled by conventional enviromental ellipsometric (EE) porosimetry\cite{boissiere_porosity_2005,bisio_interaction_2010} and by gas adsorption measurements.\cite{mooney_adsorption_1952,foo_insights_2010}. Nevertheless information retrieval from these techniques is not straightforward. {EE requires surfaces and interfaces of good optical quality and, eventually, exploitation of an index-matching fluid at the interface between the granular film and the supporting substrate (an issue in the frame of porosimetry measurements, where no other liquids should be present other than the one under investigation). For the case of EE the number of free fitting parameters is rather vast \cite{bisio_interaction_2010,toccafondi2014optical,cavaliere_exploring_2017} and, in several instances, the results may be puzzling to interpret\cite{jungk_possibilities_1993}. For instance, when dealing with metal nano-objects, as in the present case, modelling of the experimental data requires taking into account size-dependent corrections related to the presence of the surface plasmon resonance \cite{cavaliere_exploring_2017}, adding to the complexity of the information retrieval process. Furthermore, EE porosimetry is based on recognition of minute, often spectrally featureless variations of the ellipsometric angles $\Psi$ and $\Delta$ upon fluid infiltration \cite{bisio2009optical,bisio_2011}. These facts complicate the recognition of fluid filling levels as opposed, for instance, to a technique exploiting the variation of a specific resonance.}
Similar difficulties are encountered in interpreting gas adsorption measurements, where the result may be biased by the specific model, among the many available ones\cite{quirk_significance_1955}, chosen to fit the adsorption isotherm.

In this work we propose a novel sensing platform aimed at investigating fluid infiltration in nanogranular coatings.  We engineer, via atomistic simulations, a gas-phase synthesised nanogranular metallic coating with open-porosity, specifically tailored for efficient ultrafast photoacoustic detection of the filling fluid.
The detection scheme relies on impulsive photo-acoustic excitation of the device mechanical breathing modes - in the 50 GHz frequency range - and time-resolved acousto-optical read-out of the frequency shift upon fluid infiltration. From an applicative stand point, the gas-phase synthesised nanogranular metallic scaffold is readily exploitable as a distributed nanogetter coating, integrating fluid sensing capabilities and viable for multi-functionality.\cite{bettini_supersonic_2017,benetti_bottom-up_2017,galvanetto2018fodis,borghi_growth_2018}.
The production technique is  \textit{per se} competitive owing to its simplicity, high throughput and flexibility.\cite{benetti_direct_2017,corbelli_highly_2011}

The proposed scheme bears great generality and can be readily deployed to include other nanoscale\cite{nasiri2015self} or mesoscale porous systems\cite{voti2015photoacoustic,de2018towards,lamastra2017diatom}.

\section{Materials and methods}
\textbf{Molecular Dynamics.} All the simulations have been performed with the LAMMPS package\cite{LAMMPS-2}.
The velocity-Verlet algorithm was used to solve the equations of motion and temperature was controlled by a Nos\'e-Hoover thermostat. The Ag-Ag interactions were described with the 12-6 Lennard-Jones potential of \citeauthor{heinz2008}, with cutoff at 8 \AA.  Further details are reported in the SI.

\textbf{Film's effective properties.} The physical properties of the granular film have been retrieved by importing the raw xyz Molecular Dynamics (MD) results into MATLAB and by dividing the 3D simulation domain in small voxels. The average pore sizes have been computed by using an ImageJ plugin (BoneJ)\cite{doube_bonej:_2010} and the method proposed by Sainto \textit{et al.} for the bones' trabeculae analysis\cite{saito_new_1994}. The effective elastic constants are retrieved by using the MATLAB numerical solver (vpasolve) and the equations provided in the next chapters. Further details are reported in SI.

\section{Results and discussion}\label{sec:second}

\begin{figure*}
	\includegraphics[width=0.95\textwidth]{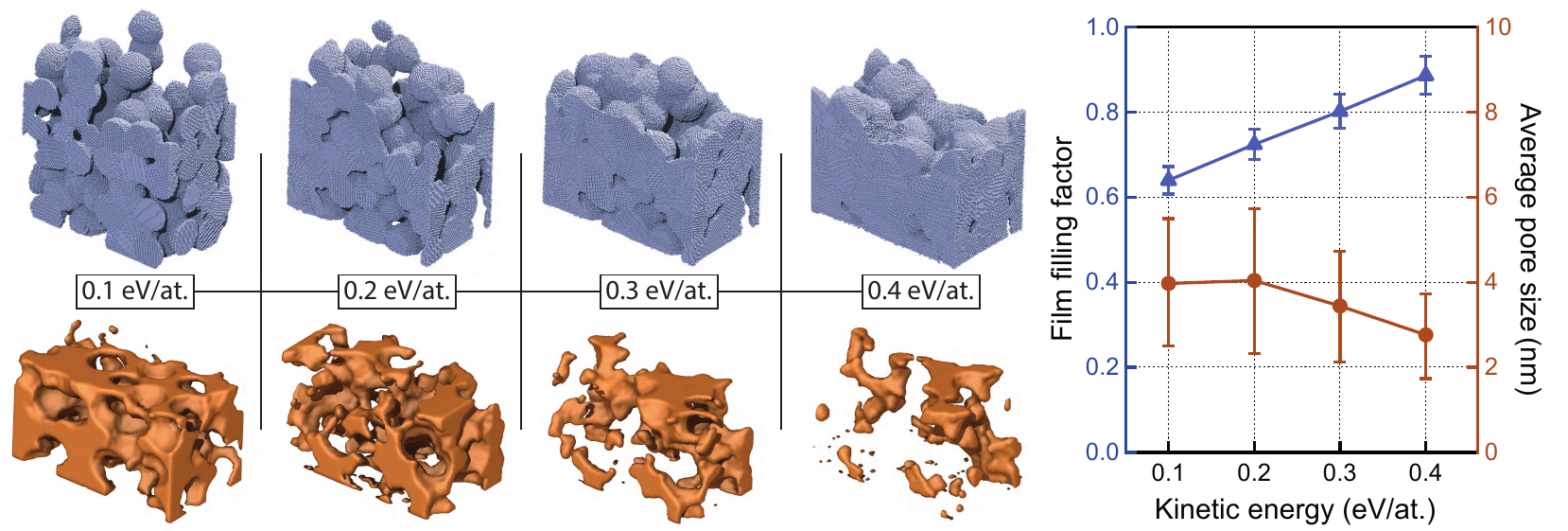}
	\caption{Left panel - top row: nanoporous scaffolds resulting from MD simulations on a 20$\times$35 nm$^2$ cell domain vs the kinetic energy per atom. Each small sphere corresponds to an atom, whereas the bigger spherical agglomerates are the NPs, partially wrapped upon landing for increasing specific kinetic energy. Left panel - bottom row: void scaffolds obtained from MD simulations vs the kinetic energy per atom. The void portion is depicted in orange colour. The renderings are the complementary of the top row ones. The latter renderings have been cut at a z-quota of 20 nm for ease of visualisation. Right panel: filling factor (blue triangles, left blue axis) and average pore dimension (red circles, right red axis) obtained from MD simulations as a function of the kinetic energy per atom of the NPs at landing. The error bars of the filling factor are due to uncertainty from MD simulations. The error bars of the average pore size are the standard deviations of the punctual pore size distribution, see SI for further details.}
	\label{Fig1}
\end{figure*}

\subsection{Device design} We consider a gas-phase synthesised Ag nanoparticles (NPs) ultrathin film deposited on polydimethylsiloxane (PDMS). It was recently demonstrated that the morphology of these films may be characterised by interconnected, channel-like pores\cite{benetti_bottom-up_2017}, thus providing a potentially wettable device. The choice of Ag, although not a stringent one\cite{bisio_2011}, is based on the availability of an experimentally validated atomistic model for a virtual film reconstruction from gas-phase deposition parameters\cite{benetti_bottom-up_2017}. Moreover, for the Ag case and for film thicknesses in the tens of nm range, these films have been experimentally proven as ultrafast photoacoustic transducers\cite{peli_mechanical_2016} with operating acoustic frequencies  spanning the range from tens to hundreds of GHz\cite{peli_mechanical_2016}. A soft polymeric support, such as PDMS, for the granular thin films yields a high acoustic film-substrate mismatch. This increases the film's breathing mode life time, thus maximising its quality factor. This fact will be further appreciated when discussing the details of the sensing scheme. Additionally, the choice of a soft polymeric support allows for a stick-on/stick-off highly bio-compatible device.
The key parameter to tune the porosity of the film is the Kinetic Energy per atom (KE) of the NPs during the deposition process. Indeed, the higher the KE, the higher the NPs deformation upon landing and, consequently, the film's filling factor. 
Assuming an average NPs diameter of 7 nm\cite{benetti_bottom-up_2017} \footnote{The actual NPs distribution is is peaked at two diameters: 1.5 nm (small NP) and 7 nm (big NP). The big NPs account for 96\% of the total deposited mass. The small NPs have been shown to be irrelevant in the reconstruction of the nanogranular thin film scaffold, its morphology, topography and mechanical properties being ruled by the big NPs only. We refer the reader to Ref. \cite{benetti_bottom-up_2017} for further details on this point.}, we implemented four different MD simulations, tuning the KE of the NPs-forming atoms across the set of values \{0.1, 0.2, 0.3, 0.4\} eV/at, the latter range being quite typical in Supersonic Cluster Beam Deposition (SCBD). In brief, each simulation reproduces the landing of 90 NPs of $\sim$ $10^{4}$ atoms on a 20x35nm$^2$ base domain, for a total of $\sim$ 0.9 million atoms. In gas-phase NP sources, such as a SCBD apparatus, the KE of the NPs can be tuned by varying the source's geometrical parameters\cite{VahediTafreshi2002} and/or carrier gas type and temperature\cite{mazza_accessing_2011}.

The rendering of the virtual films, resulting from MD simulations, is reported against KE in Figure \ref{Fig1}, left panel - top row. The virtual films are composed of NPs (spherical agglomerates in Figure \ref{Fig1} left panel - top row). Every NP is assembled atom-by-atom (the atoms are the smaller spheres visible upon adopting an high magnification for figure inspection).
The void scaffolds are reported for increasing KE in Figure \ref{Fig1}, left panel - bottom row, the void portion being depicted in orange. The void scaffolds are the complementary of the NPs film scaffolds.
For each scaffold, the film filling factor, $\mathit{FF} = V_{NP}/V$ where $V_{NP}$ is the total volume of all the NPs and $V$ is the overall film volume\cite{benetti_bottom-up_2017}, and the average pore size\cite{saito_new_1994,JMI:JMI134,doube_bonej:_2010} are reported against KE in Figure \ref{Fig1}, right panel, {refer to SI for further details}. Simulations results show that, decreasing the KE from 0.4 to 0.1 eV/at, the filling factor decreases from 0.9 to 0.64 whereas the void morphology evolves from sparse, mostly clogged pores to an open-pore trabecular-like structure. The average pore size increases from 2.8 to 4 nm and the average film thickness $h$ from 25 to 35 nm.

Among the simulated granular films, the best geometrical features for device engineering are achieved for a deposition KE of 0.1eV/at. Indeed, the connected pores morphology, together with the lower $\mathit{FF}$=0.64 and the biggest pores of this film - 4 nm average pore diameter - yield maximum sensitivity and higher storage capacity while minimising the pinning-related issues, which impede fluid infiltration inside the film. Finally, a value of KE = 0.1 eV/at. should prevent in-depth NPs implantation in PDMS, as is expected for instance for KE=0.5 eV/at.\cite{ravagnan_polymethyl_2009}.
We will henceforth contextualise the discussion focusing on the Ag granular ultrathin film of thickness $h$=35 nm obtained with KE = 0.1 eV/at. 

\begin{figure}
	\includegraphics[width=0.49\textwidth]{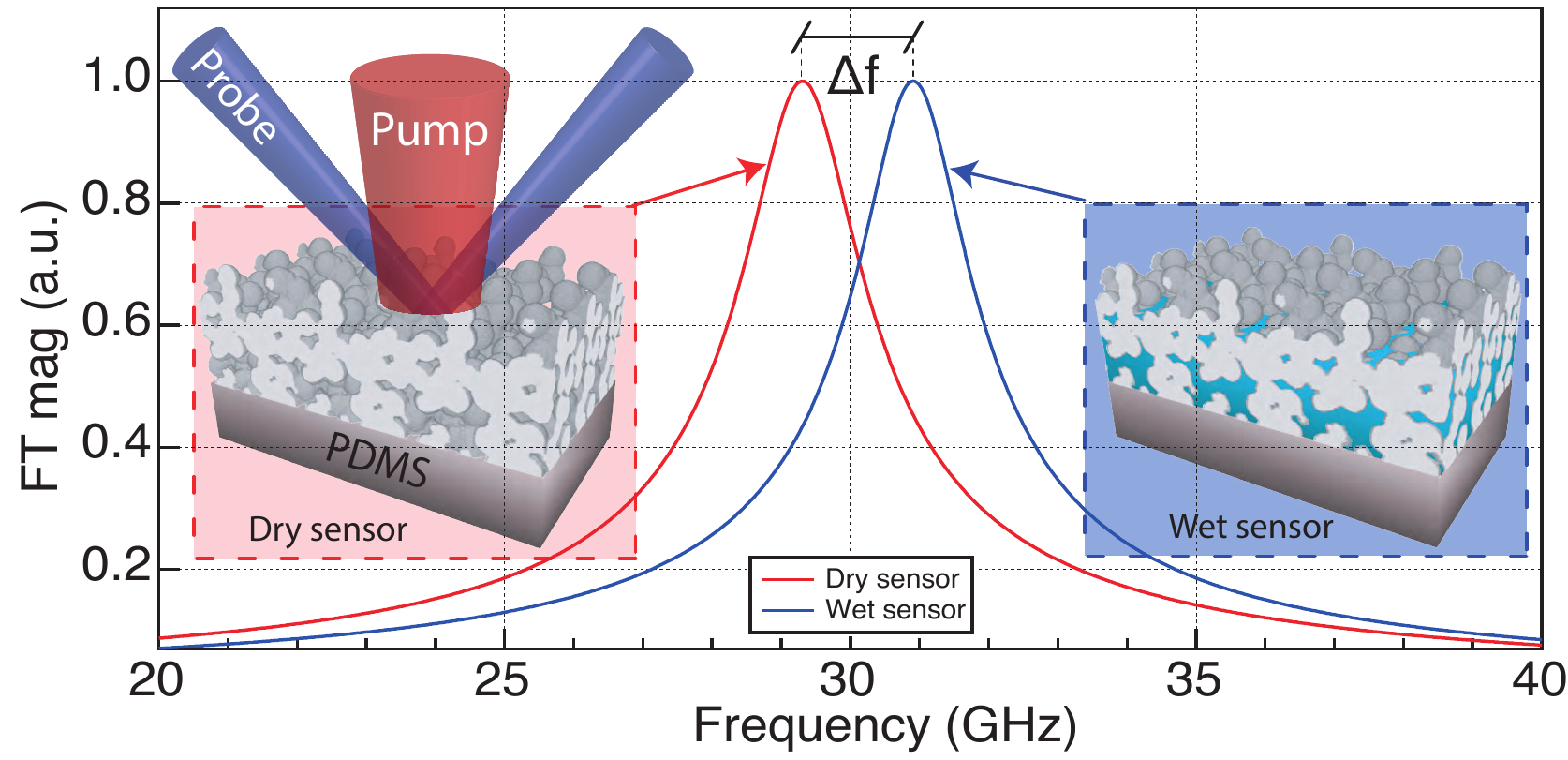}
	\caption{Fourier transform magnitude $|\tilde{F}(f)|$ of the photoacoustic signal expected for a dry (red) and fully infiltrated (blue) granular thin film sensor. The resonances are ascribed to the fundamental, n=1, acoustic breathing modes. Upon water infiltration the acoustic resonance undergoes a frequency shift $\Delta f$=$f_{wet}$-$f_{dry}$ together with a linewidth reduction from $\Gamma_{dry}$ to $\Gamma_{wet}$. 
The two insets represent the dry and fully infiltrated device respectively, together with a schematics of the pump and probe technique.}
	\label{Fig2}
\end{figure}

\subsection{Sensing scheme} The sensor working principle relies on ultrafast photoacousics\cite{matsuda2015fundamentals}, among the emerging techniques for mechanical nanometrology in a variety of systems ranging from phononic crystals\cite{travagliati2015,nardi2012design,mante_probing_2015}, ultrathin-films\cite{nardi2015,hoogeboom2016,grossmann2017}, to multilayer tube\cite{brick2017} and granular materials\cite{avice2017}. The transduction scheme is as follows: in the excitation step (opto-acoustic transduction) an ultrafast IR laser pump pulse illuminates the device. {The energy absorbed by the granular metallic film leads to an impulsive lattice temperature increase of the order of a few Kelvins, avoiding any annealing effect.}\footnote{This value is obtained accounting for the typical fluences and repetition rates in use in Fiber-Laser \cite{peli_mechanical_2016} or Ti:sapphire oscillator based set-ups \cite{PhysRevB.95.085306}.}. The subsequent thermal expansion excites the film's breathing modes at their frequencies $f_{n}$, $n$ being the mode order. The excited breathing modes decay with time-constants $\tau_n$, radiating acoustic waves into the substrate. In the detection step the excited breathing modes modulate the film's dielectric constants (acousto-optic transduction). The acoustic oscillations are ultimately encoded in the relative reflectivity/transmissivity variations, $\Delta$$I$/$I_{0}$, as measured by a time-delayed probe pulse, the time-delay being with respect to the instant of the pump-pulse arrival $t$=0\cite{peli_mechanical_2016}. Here $\Delta$$I$=$I(t)$-$I_{0}$, where $I(t)$ is the reflectivity/transmissivity at time-delay $t$ and $I_{0}$ the reflectivity/transmissivity of the unexcited sample (static reflectivity/transmissivity). The contribution of the breathing mode $n$ to the time-resolved optical trace, once thermal effects are filtered out, is well mimicked by:
\begin{equation}
\label{sign}
 F_{n}(t)\equiv(\Delta I/I)_{n}= A_n  e^{-t/\tau_n} \, cos(2\pi f_n \, t + \phi_n) \, \theta(t)
\end{equation}
where $\theta(t)$ is the Heaviside step function, $\tau_n$, $f_n$, $\phi_n$ and $A_{n}$ the breathing mode life-time, frequency, phase and amplitude contribution to the optical signal respectively. Calculating the magnitude of the Fourier transform (FT) of the time resolved trace given by Eq. \ref{sign}, under the assumption $f_{n}>>1/\tau_{n}$, we obtain the following resonance line shape:
\begin{equation}
\label{FTsign}
|\tilde{F_{n}}(f)| = \frac{|A_n| \ \tau_n}{2\sqrt{1+[2 \pi \tau_n (|f|-f_n)]^2}} 
\end{equation}
We refer the reader to SI for further details on this point.

The quantities ruling the resonance, and of relevance to the present discussion, are $f_{n}$, $\tau_{n}$ and their interplay synthesised by the quality factor $Q_{n}$. With this notion in mind, we now inspect the device acoustic properties within the frame of the effective medium approximation (EMA). The granular NPs thin film, weather fluid-infiltrated or not, is modelled as an \textit{effective} homogeneous and isotropic thin film of $\textit{effective}$ density, $\rho^*$, and $\textit{effective}$ stiffness tensor, $\mathbf{C}^*$. The longitudinal sound velocity is $v_{NP}= \sqrt{C^*_{11} /\rho^*}$, where $C^*_{11}$ is the first element of the film effective stiffness tensor. The acoustic impedance reads $Z = \sqrt{C^*_{11} \, \rho^*}$. The granular film adheres on a PDMS substrate of acoustic impedance $Z_S<Z$. Within this frame, the relevant breathing mode parameters read:
\begin{equation}
\label{fhomo}
f_n = \frac{v_{NP}}{2 h} n = f_1 n 
\end{equation}

\begin{equation}
\label{tauhomo}
\tau_n = \left| f_1 ln \left( \frac{Z-Z_S}{Z+Z_S} \right) \right| ^{-1} 
\end{equation}

\begin{equation}
\label{qhomo}
Q_n = \pi f_n \tau_n = \pi \left| ln \left( \frac{Z-Z_S}{Z+Z_S} \right) \right|  ^{-1} n
\end{equation}
We refer the reader to SI for further details on the model and equations derivation.

A ``dry" device is characterised by a resonance centred at a frequency $f_{dry}$ and with decay-time $\tau_{dry}$ (we dropped the mode index $n$ for brevity). Upon fluid infiltration in the granular film, both $\rho^*$ and $C_{11}^*$ increase, leading to a resonance of frequency $f_{wet}$ and lifetime $\tau_{wet}$. 
For the sake of exemplification we here anticipate results that will be derived further on. Figure \ref{Fig2} reports the FT modulus of the fundamental breathing mode, n=1, expected for the dry (red resonance) and for the fully water infiltrated - fully wet - device (blue resonance). Upon full water filling the resonance shifts by an amount $\Delta$$f$=$f_{wet}$-$f_{dry}$=1.61 GHz and the linewidth decreases from $\Gamma_{dry}$= 2.83 GHz to $\Gamma_{wet}$=2.68 GHz, where $\Gamma_n = \sqrt3 / ( \pi \tau_n)$ is the resonance's FWHM. The frequency shift and the decay time variations thus allow to quantify the amount of infiltrated fluid. 
{The resonance frequency is first measured on the bare device (dry configuration). The device is then loaded (wet configuration) and its resonance frequency measured. The resonance frequencies of the dry and wet device are hence acquired in separate measurement sessions.}\footnote{This is at variance with respect to the problem of separating two peaks from a signal which is the superposition of them, such as resolving the two diffraction peaks in a double slit optical diffraction experiment.}. The minimum resolvable shift has thus to be considered as the error in the estimation of the peaks centers. The higher the resonance Q-factor, the smaller is the error in the peak center estimation, a high Q-factor thus being a desirable feature.
\subsection{The Practical Case} In general, the detection strategy is based on resolving the resonance frequencies between the wet and dry configurations, linking $\Delta f$ to the amount of filling fluid. We here illustrate the strategy for the paradigmatic case of water adsorption, the idea being alike for other fluids. We chose water owing to its relevance in biology and bio-related applications. We consider the optimised device, \textit{i.e.}, the one obtained with KE of 0.1eV/at. 
{The interconnected porous structure and pores size allow for a homogeneous water distribution in the whole accessible volume, thus justifying an EMA approach (see SI for further information on this point); other infiltration scenarios will be addressed further on.}
The device is therefore considered as an effective homogeneous and isotropic film of effective density, $\rho^*$, effective shear, $G^*$, and bulk, $K^*$, modulus (the effective stiffness tensor $\mathbf{C}^*$ being completely defined by $G^*$ and $K^*$ within the EMA). These quantities depend of the amount of infiltrated water. The density $\rho^*$ is obtained as a weighted mean of the silver and water densities ($\rho_{Ag}$ and $\rho_{w}$ respectively) on the corresponding occupied volume in the scaffold: $\rho^*(l) = \rho_{Ag}\, \mathit{FF} + \rho_{w} (1-\mathit{FF}) \, l$. The relative loading, $l = V_{w}/ V_{void}$, is the infiltrated water volume, $V_{w} $, normalized against the total volume available for infiltration, $V_{void}$. $G^*$ and $K^*$ are retrieved numerically solving Budiansky equations \cite{budiansky_elastic_1965}:
\begin{equation}
\label{budi}
\begin{aligned}
 \sum_{i=1}^{N} \frac{c_i}{1+ \frac{3K^*}{3K^* +4G^*}  \left( \frac{K_i}{K^*} -1 \right) } =1 \\
 \sum_{i=1}^{N} \frac{c_i}{1+\frac{6(K^*+2G^*)}{5(3K^*+4G^*)} \left( \frac{G_i}{G^*} -1 \right) }=1
 \end{aligned}
\end{equation}
where the index $i$ runs over the $N=3$ materials composing the effective film (i = Ag, water, voids); $K_i$ and $G_i$ are the constituents bulk and shear modulus; $c_i$ are the constituents normalised concentrations: $c_{Ag} = \mathit{FF}$, $c_{water}= (1-\mathit{FF})\, l$ and $c_{voids}= (1-\mathit{FF})\, (1-l)$. We stress that $K^*=K^*(l)$ and $G^*=G^*(l)$, are both functions of the loading $l$. This dependence has been omitted in Eq. \ref{budi} for ease of visualisation. $C^*_{11}$ is a function of $l$ through the equality $C^*_{11} = K^* + 4 G^* / 3$. All the above-mentioned materials parameters are reported in SI. We are therefore in the position to calculate $f_{n}$, $\tau_{n}$ and $Q_{n}$ of the first two breathing modes, $n$=1 and $n$=2, as a function of the relative loading $l$ (see Equations \ref{fhomo}, \ref{tauhomo} and \ref{qhomo}). 
\begin{figure}[t]
	\includegraphics[width=0.45\textwidth]{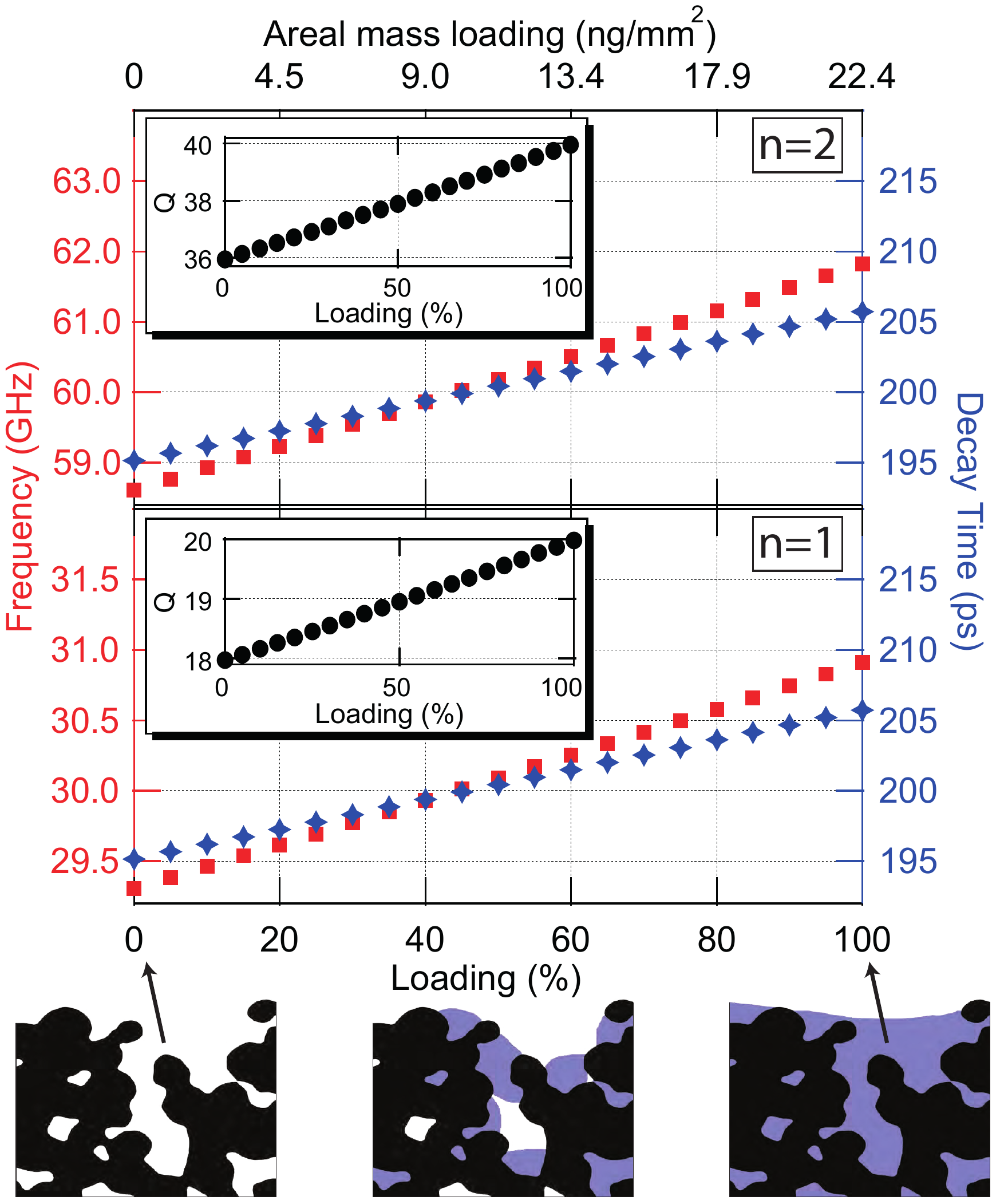}
	\caption{Frequency of the n=1 (bottom panel) and n=2 (top panel) acoustic breathing modes (red squares, left red axis) and decay times (blue diamonds, right blue axis) vs water filling within the homogeneous adsorption scenario. Water filling is expressed both as relative volumetric loading $l$ (bottom axis) and equivalent areal mass loading $m_S$ (top axis).
Insets: quality factor $Q$ vs relative volumetric loading $l$. Cartoon: schematics of the infiltrated device for $l$=0, 50\% and 100\%. Water is depicted in blue and silver in black.}
	\label{Fig3}
\end{figure}

The results are reported in Figure \ref{Fig3}. The frequencies $f_{n}(l)$ are linear with $l$ (red squares, left vs bottom axis), so as the frequency shifts, $\Delta f_{n}(l)$=$f_{n}(l)$-$f_{n}(0)$. The decay times $\tau_{n}$ (blue diamonds) are mode-independent and, consistently, $Q_{2}=2Q_{1}$ (see insets of Figure \ref{Fig3}). The high value of $Q$, as compared to the values recently reported on similar systems\cite{peli_mechanical_2016}, $Q \approx 1-5$, stems from the tailored choice of the substrate material. A soft substrate maximises the acoustic impedance mismatch, the device approaching the ideal free-standing case. 

We now focus on the device sensitivity issue. {Making the necessary changes} from Ref. \citenum{vellekoop1998acoustic}, we introduce $S_{l} \equiv \frac{df_n(l)}{dl} \frac{1}{f_n(l)}$ as a sound definition to quantify the device's sensitivity to liquid infiltration. Provided the linearity of $f(l)$ (see Figure \ref{Fig3}) and that $\Delta f_{n}(l=1) \ll f_{n}(l)$, the sensitivity reads $S_{l}\approx \frac{\Delta f(l=1)}{1}\frac{1}{f(0)} = 0.05$. Furthermore, the sensitivity $S_{l}$ is thickness- and mode-independent: from Eq. \ref{fhomo} $S_{l} = \frac{v_{NP}(l=1)-v_{NP}(l=0)}{v_{NP}(l=0)}$ where no dependence on $h$ and $n$ appears. On the contrary $Q$ is mode-dependent, as shown in the insets of Figure \ref{Fig3}. As previously mentioned, a high $Q$ is a desirable feature to minimise the error in frequency shift read-out, thus privileging higher modes for sensing purposes.
\\
For the sake of comparison against typical figures of merit\cite{vellekoop1998acoustic, nardi2012design}, we introduce the areal mass loading $\mathit{m_S}= c_{water}\, h\,  \rho_w = (1- \mathit{FF})\, l\, h\,  \rho_w$. The value $\mathit{m_S}$ quantifies the mass of infiltrated water normalized on the device's unit area. The standard sensitivity to mass-loading definition reads\cite{vellekoop1998acoustic} $S_{m_s} \equiv \frac{df(m_S)}{dm_S} \frac{1}{f(m_S)}$. Holding the same approximations discussed to evaluate $S_{l}$, the mass sensitivity reads $S_{m_s} \approx \frac{\partial f}{\partial l}\frac{\partial l}{\partial m}\frac{1}{f(0)}=S_{l}\frac{l}{(1-FF) \rho_{w} h} = 26 \times 10^{3}$ cm\textsuperscript{2}/g. {The latter figure can be further increased, since it scales as $1/h$ ($S_{l}$ is h-independent). A minimum value of h in excess of 14 nm is a realistic figure, granting a film with a fully developed granularity \cite{benetti_bottom-up_2017}.
\\
For the sake of comparison, we note that the proposed device sensitivity outperforms, by three orders of magnitude, that of commercially available quartz crystal microbalances (QCM) and, by an order of magnitude, that of flexural plate wave (FPW) devices, see Table \ref{compSens}.}
\begin{table}[t]
	\newcolumntype{x}[1]{>{\centering\arraybackslash\hspace{0pt}}p{#1}}
	\begin{center}	
		\begin{tabular}{m{2.5cm} c x{2cm}}
			\hline			Device type 		&	$f_0$ (GHz) & $S_{m_s}$\\\hline
			NOT & $3\times 10^{1} $	& $26 \times 10^{3}$	\\
			QCM & $6.0\times 10^{-3}$ 	& 14	\\
			FPW & $2.6\times 10^{-3}$	& 951	\\\hline
		\end{tabular}
	\end{center}
	\caption{{Comparison of operation frequencies and mass sensors�� sensitivities. NOT stands for nanogranular optoacoustic transducer (this work), QCM for quartz crystal microbalance and FPW for flexural plate wave devices. The values for typical QCM and FPW are taken from ref \cite{cheeke1999acoustic}.}}
	\label{compSens}
\end{table}

The frequency vs loading curves are well within the detectability range of current ultrafast photo-acoustic technology. In Figure \ref{Fig3} we calculated the device response discretising the liquid loading in steps of 5$\%$, resulting in frequencies separated by $\sim$ 0.1 GHz for the n=1 case (0.2 GHz for n=2).  Nevertheless, the frequency resolution which may be achieved with state-of-the art photo-acoustic technology is way higher. For instance, subharmonic resonant optical excitation of acoustic modes in thin membranes allows resolving frequency shifts with megahertz resolution.\cite{bruchhausen2011}. The minimum detectable infiltrated liquid variation reads $dl = \frac{df}{f(0)}\frac{1}{S_{l}}$. Taking $df \sim$ 1 MHz, $S_{l}\sim$0.05 and $f(0)=f_{n=1}(0)\sim$29.5 GHz for mode $n$=1 we find $dl\sim$7$\times10^{-4}$. This value can be further decreased exploiting higher modes, in fact $dl$ scales as $\frac{1}{f_{n}(0)}\sim\frac{1}{nf_{0}(0)}$. This fact may be readably appreciated comparing the slopes of the frequency vs loading curves for the $n=1$ and $n=2$ reported in Figure \ref{Fig3}. 
\\
{
\indent As opposed to Inter Digital Transducer (IDT) technology, within the present sensing scheme acoustic wave generation, detection and sensing of the infiltrated fluid take place in the same active region, identified by the probe beam spot size. This fact favours miniaturization and allows working with minute quantities of total infiltrated fluid. For instance, assuming a typical probe beam diameter of 10 $\mu$m a fully infiltrated active area is tantamount to $\sim$1 femtoliter of infiltrated fluid.}

\subsection{Discriminating among different infiltration patterns} 
\begin{figure*}[t]
	\includegraphics[width=0.95\textwidth]{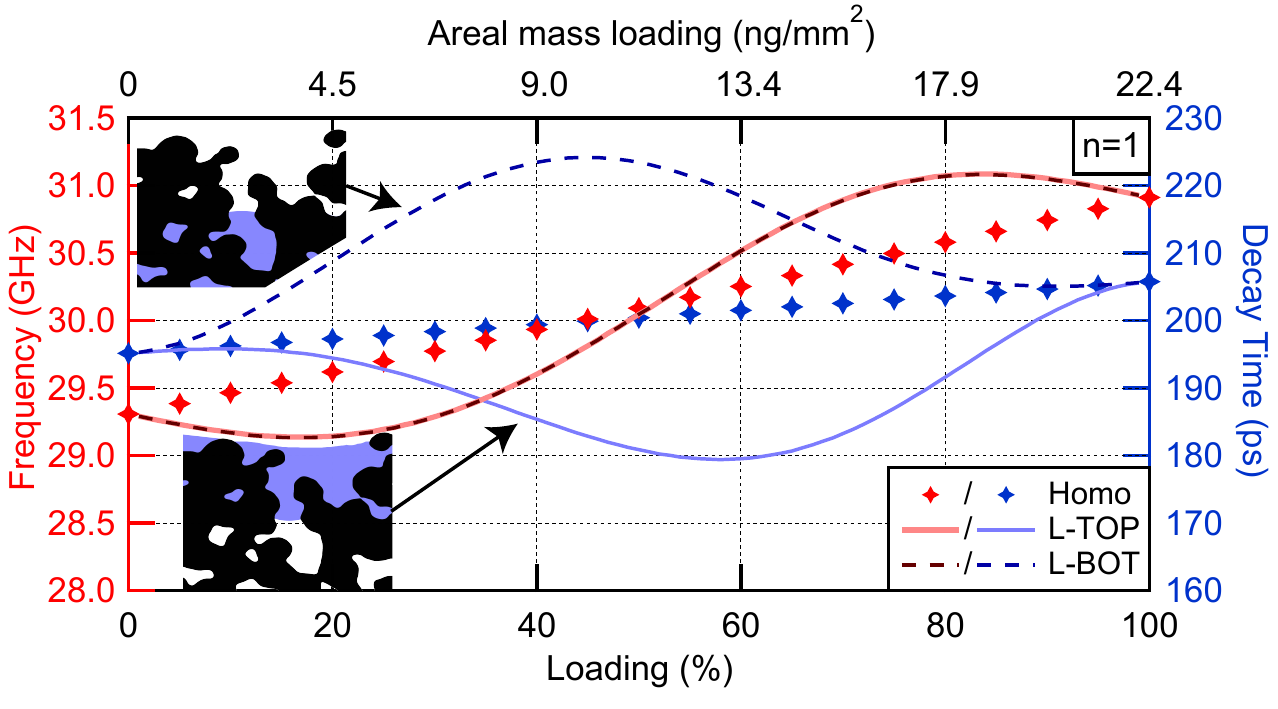}
	\caption{Frequency of the n=1 acoustic breathing modes (left axis, red) and decay times (right axis, blue) vs water filling within the layered adsorption scenarios L-TOP (full lines), L-BOT (dashed lines) and, for sake of comparison, for the homogeneous wetting case (markers). Water filling is expressed both as relative volumetric loading $l$ (bottom axis) and equivalent areal mass loading $m_S$ (top axis). Cartoon: schematics of the infiltrated device for the L-TOP (bottom cartoon) and L-BOT (top cartoon) scenarios. Water is depicted in blue and silver in black.}
	\label{Fig4}
\end{figure*}

In the previous discussion we assumed a homogeneous infiltration scheme. We now argue that the proposed sensing platform may be exploited to gain insight into the fluid infiltration pattern, a yet unsolved issue in nanoporous materials and forming the object of extensive research.\cite{boissiere_porosity_2005,Huang2017,zhang2017precise} 
Let's analyse the layered infiltration pattern which comprises two scenarios. The ``water layer on top" (L-TOP) scenario consists of a water infiltrated layer sitting on top of an empty one (see inset at bottom-left corner of Figure \ref{Fig4}).
{This situation might be expected for high filling factors, such as the one predicted for KE of 0.4eV/at. In this situation some inner pores might be clogged and not accessible, as suggested by Bisio \textit{et al.} \cite{bisio_interaction_2010}.
The opposite scenario, ``water layer on bottom" (L-BOT), consists in a water infiltrated layer sitting on the polymeric substrate and covered by an empty layer on top (see inset in top-left corner of Figure \ref{Fig4}). This scenario might arise, for instance, when water starts evaporating from a fully infiltrated scaffold. In this context one may for instance follow the evaporation process as it takes place, tracking in time the loading curve evolution. We pinpoint that, for the case of a fully infiltrated device, the L-TOP and L-BOT scenario coincide.}

The stratified scenario is conveniently modelled as two effective media in series deposited on a semi-infinite substrate, the effective media mimicking the \textit{fully} wet (fully water infiltrated-Ag scaffold: Ag and water filled voids) and dry layers (non-infiltrated Ag scaffold: Ag and empty voids). For the layered-case the breathing mode frequency and decay time read $f_{n}$=Re$\{\omega_{n}\}$/2$\pi$ and $\tau_{n}$=1/$\lvert$Im$\{\omega_{n}\}$$\rvert$ respectively, where the complex-valued $\omega_{n}$ is the $n^{th}$ root of the following equation \cite{peli_mechanical_2016}:
\begin{equation}
\label{eqw}
r_{t,b} \ e^{-2i\omega \frac{h_t}{v_t} } + r_{b,s} e^{-2i\omega (\frac{h_t}{v_t} + \frac{h_b}{v_b})} - r_{t,b} \, r_{b,s} e^{-2i\omega \frac{h_b}{v_b}} -1 =0.
\end{equation}
The subscripts \textit{t}, \textit{b} and \textit{s} stand for top layer, bottom layer and substrate respectively; $r_{m,n} = (Z_{m}-Z_n)/(Z_m+Z_n)$ is the acoustic reflection amplitude between the $m$ and $n$-indexed layers and $Z_n$, $v_{n}$ and $h_n$ the $n$-indexed layer acoustic impedance, longitudinal sound velocity and thickness respectively. $C^*_{11}$ for the \textit{fully} dry and wet layers (corresponding to the homogeneous wetting case of l=0 and l=100\% respectively) are calculated from Equation \ref{budi}, whereas $\rho^*$ equals $\rho_{Ag}$ and $\rho_{Ag}\, \mathit{FF} + \rho_{w} (1-\mathit{FF})$ for the dry and wet layers respectively. These quantities allow calculating $Z_b$, $Z_t$ and $v_{b}$, $v_{t}$. 
The total film thickness is $h=h_t+h_b = 35 \ \mathrm{nm}$. The relative water loading reads $l$=$h_{t}$/$h$ for the L-TOP case ($l$=$h_{b}$/$h$ for L-BOT) .
For sake of comparison, we analyse the acoustic response keeping the same scaffold as in the homogeneous wetting case, i.e. the one obtained for KE = 0.1 eV/at.
The breathing mode frequency, $f_{1}$, and decay time, $\tau_{1}$, for the n=1 mode, obtained upon numerical solution of Equation \ref{eqw}, are reported in Figure \ref{Fig4}. As opposed to the homogeneous wetting case, both $f_{1}$ and $\tau_{1}$ are non linear vs $l$.

\noindent The L-TOP and L-BOT are characterised by the same $f_{1}$ vs loading curve (continous and dashed red lines respectively in Figure \ref{Fig4}). This is due to the fact that, with regards to $f$, the boundary condition at the nanoporous film-PDMS interface is substantially stress free. With respect to $f$ the device behaves as a free-standing layered membrane, resulting in a symmetric situation between the L-TOP and L-BOT scenarios.

\noindent Conversely, the $\tau_{1}$ vs loading curves (continous and dashed blue lines in Figure \ref{Fig4}) show opposite trends in the L-TOP and L-BOT scenarios. In particular, whatever the quantity of adsorbed water, $\tau_{1}(l)$ is smallest for the L-TOP scenario, is biggest for the L-BOT scenario and sits between the two for the homogeneous filling case. The physical explanation has to do with the acoustic impedance jumps across the device thickness. The smoother are the impedance changes across the device, the highest is the acoustic wave transmission to the PDMS substrate, hence the lower is the breathing mode damping time. In the L-TOP scenario the acoustic impedance decreases across the sample from the device free-surface all the way into the PDMS layer: $Z_t >Z_b>Z_s$. This scenario maximises acoustic transmission from the device to the substrate, hence leading to the lowest decay times. The L-BOT scenario is characterised by the greatest acoustic impedance jumps across the sample: $Z_t <Z_b$ and $Z_b>Z_s$, hence leading to the longest decay times. The homogeneous filling case lays in between: $Z_{Homo}>Z_s$ where $Z_{Homo}$, the nanoporous film acoustic impedance for the homogeneous water infiltration case, 
ranges between the fully wet and dry device, corresponding to l=0 and l=100\% respectively.
The cases for n=2 and n=3 are discussed in SI.

The $f(l)$ curves thus allow discriminating between homogeneous and layered infiltration scenarios. As for the latter, the L-TOP and L-BOT cases may be differentiated on the basis of the $\tau(l)$ (or $Q(l)$) curves. This proves the potential of the present strategy in uncovering the fluid infiltration pattern in nanoporous materials.

\section{Conclusions}

We designed a novel sensing platform allowing to quantify the amount of fluid infiltrated in a nanoporous coating.
The platform is based on a gas-phase synthesised nanogranular metallic coating with open-porosity, specifically engineered via molecular-dynamics for efficient ultrafast photoacoustic detection of the filling fluid. For the paradigmatic case of water filling we predict a sensitivity exceeding 26$\times$10$^{3}$ cm\textsuperscript{2}/g, upon equivalent areal mass loading of few ng/mm\textsuperscript{2}, outperforming current microbalance-based technology by three orders of magnitude. The predictions are robust, the theoretical frame having been recently benchmarked against experiments\cite{benetti_bottom-up_2017}, and ultrafatst photoacoustic read-out of mechanical resonances in metallic nanoporous coatings demonstrated\cite{peli_mechanical_2016}.

The nanogranular metallic scaffold is readily exploitable as a distributed nanogetter coating, integrating fluid sensing capabilities and serving as a stick-on/stick-off highly bio-compatible device. The film may be deposited on virtually any surface while varying the metal composition\cite{benetti_direct_2017,corbelli_highly_2011}, allowing to foresee integration of the fluid sensing capabilities with a variety of applications\cite{bettini_supersonic_2017,benetti_bottom-up_2017,borghi_growth_2018}. The present strategy, relying on the variation of a specific acoustic resonance upon fluid infiltration, provides data that are intrinsically simpler to interpret as compared to existing techniques such as EE and gas adsorption-based porosimetry. For this reason it might serve as a valid alternative/complement to current technologies.

Finally, we showed that the proposed sensing scheme allows discriminating among different filling patterns, providing a means to investigate pinning-related issues of relevance for nanoporous membrane wettability.

The present scheme is general and easily implementable. It may be expanded to other emerging gas-phase synthesized granular materials\cite{nasiri2015self} and extended to materials with mesoscale porosity\cite{voti2015photoacoustic,de2018towards,lamastra2017diatom}. {In perspective, the present scheme could be expanded to include granular multi-layers synthesized with different metals. This would allow expanding the range of exploitable acoustic parameters so as to fine-tune the loading curve when choosing other substrate materials, for instance by engineering an acoustic-impedance graded device. Furthermore, this strategy would allow to exploit a Ti granular film as an adhesion layer to increase the overall device sticking-factor when working with rigid substrates.} The sensing apparatus may be implemented taking advantage of readily available compact, table-top sources relying on superior sampling speed and telecom technology - such as the ASOPS technique\cite{nardi2015,schubert2012spatial} - and EUV coherent sources granting superior photoacoustic sensitivity\cite{nardi2013probing,nardi2011probing}.

\section{Supporting Information}
Device design details (MD details, Filling factor, Volume and thickness, pore size), Sensing scheme (Derivation of Eq.2 and its applicability,  Details of the model: derivation of Eq.s 3 and 4, A possible mechanism to convey water to the device), The Practical Case (Materials constant, Layered cases for n$\leq$3).

\section{Acknowledgements}
G.B. acknowledges financial support from the Research Foundation Flanders (FWO). C.C. and F.B. acknowledge financial support from the MIUR Futuro in ricerca 2013 Grant in the frame of the ULTRANANO Project (project number: RBFR13NEA4). F.B. and C.G. acknowledge support from Universit\`a Cattolica del Sacro Cuore through D.2.2 and D.3.1 grants. F.B. acknowledge financial support from Fondazione E.U.L.O.
C.C. acknowledges financial support from Programma Operativo Nazionale 2007-2013 ``Ricerca e competitivit\`a" financed by EU through project PON04a2\_00490 ``Ricerca Applicata a Reti di comunicazione M2M e modem integrati innovativi dedicati a servizi avanzati per le Smart Cities - M2M Netergit", and computational support from PRACE for awarding access to Marconi hosted at CINECA, Italy, through project UNWRAP (call 14), and ISCRA through project UNWRAPIT.
\bibliography{G3_refs}


\end{document}